\documentclass{article}
\usepackage{spconf,amsmath,graphicx}
\usepackage{url}
\usepackage{amssymb}
\usepackage{multirow}
\usepackage{caption}
\usepackage{subcaption}
\usepackage{color}
\usepackage{tikz}
\usepackage{booktabs}
\usepackage{enumitem}
\usepackage{multirow}
\usepackage{multicol}
\usepackage{pifont}
\usepackage{hyperref}
\usepackage{tablefootnote}

\title{HTS-AT: A hierarchical token-semantic audio transformer \\ for sound classification and detection }
\name{Ke Chen $^1$, Xingjian Du $^2$, Bilei Zhu $^2$, Zejun Ma $^2$, Taylor Berg-Kirkpatrick $^1$, Shlomo Dubnov $^1$}
\address{
         $^1$University of California San Diego \quad
	     $^2$AI Lab, Bytedance Inc.
	     }

\begin{document}
%
\maketitle

\begin{abstract}
Audio classification is an important task of mapping audio samples into their corresponding labels. 
Recently, the transformer model with self-attention mechanisms has been adopted in this field. 
However, existing audio transformers require large GPU memories and long training time, meanwhile relying on pretrained vision models to achieve high performance, which limits the model's scalability in audio tasks. 
To combat these problems, we introduce HTS-AT: an audio transformer with a hierarchical structure to reduce the model size and training time. It is further combined with a token-semantic module to map final outputs into class featuremaps, thus enabling the model for the audio event detection (i.e. localization in time). 
We evaluate HTS-AT on three datasets of audio classification where it achieves new state-of-the-art (SOTA) results on AudioSet and ESC-50, and equals the SOTA on Speech Command V2. It also achieves better performance in event localization than the previous CNN-based models. 
Moreover, HTS-AT requires only 35\% model parameters and 15\% training time of the previous audio transformer. These results demonstrate the high performance and high efficiency of HTS-AT.

\end{abstract}
\begin{keywords}
Audio Classification, Sound Event Detection, Transformer, Token-Semantic Module
\end{keywords}

\section{Introduction}
\vspace{-0.2cm}
Audio classification is an audio retrieval task which aims to learn a mapping from audio samples to their corresponding labels. Depending on the audio categories, it involves sound event detection \cite{sed-tutor}, music instrument classification \cite{cmg-ac}, among others. It establishes a foundation for many downstream applications including music recommendation \cite{ke-recom}, keyword spotting \cite{speechcommandsv2}, music generation \cite{ke-sketchnet, haowen-muspy}, etc.

With burgeoning research in the field of artificial intelligence, we have seen significant promising progress in audio classification. 
For data collections, many datasets with different types of audio (e.g. AudioSet \cite{audioset}, ESC-50 \cite{esc50}, Speech Command \cite{speechcommandsv2}, etc.) provide platforms for the training and evaluation of models on different subtasks. 
For the model design, the audio classification task is thriving based on neural-network-based models. 
Convolutional neural networks (CNNs) have been widely used in this field, such as DeepResNet \cite{deepres}, TALNet \cite{talnet}, PANN \cite{pann}, and PSLA \cite{psla}. These models leverage CNN to capture features on the audio spectrogram, and further improve their performance through the design of the depth and breadth of the network. 
Recently, by introducing the transformer structure \cite{transformer} into audio classification, the audio spectrogram transformer (AST) \cite{ast} further achieves the best performance through the self-attention mechanism and the pretrained model from computer vision. In this paper, we take a further step on a transformer-based audio classification model by first analyzing remaining problems in the AST.

\begin{figure*}
    \centering
    \includegraphics[width=\textwidth]{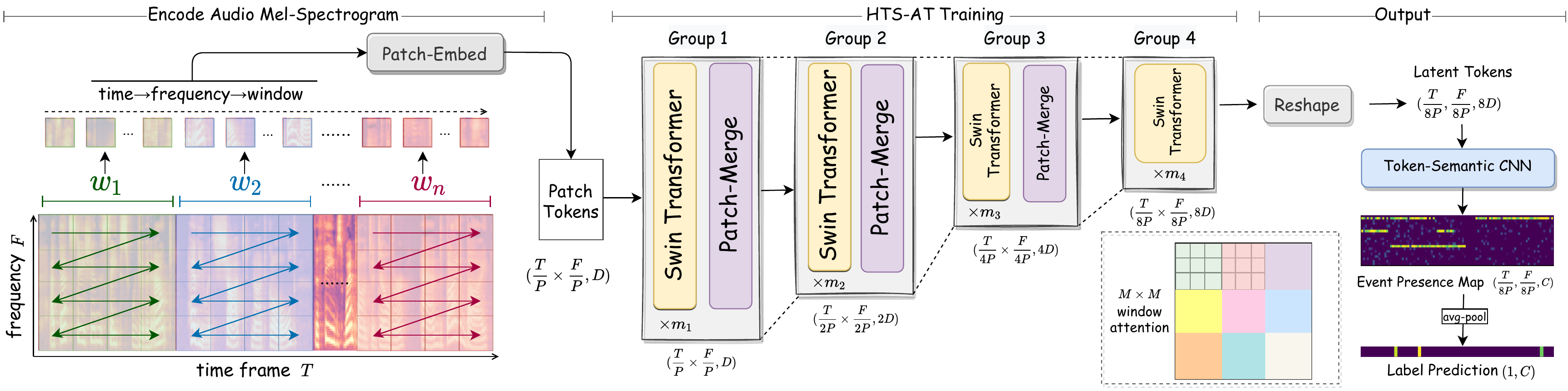}
    \vspace{-0.6cm}
    \caption{The model architecture of HTS-AT.}
    \label{fig:model-arch}
\vspace{-0.5cm}
\end{figure*}

First, since the transformer takes the audio spectrogram as a complete sequential data, AST takes a long time to train and consumes large GPU memories. In practice, it takes about one week to train on the full AudioSet with four 12GB GPUs. 
One method to boost training speed is to use the ImageNet \cite{imagenet} pretrained model in computer vision. However, this also limits the model to those pretrained hyperparameters, which reduces its scalability in more audio tasks. 
Indeed, we find that without pretraining, AST can only achieve the baseline performance (mAP=0.366 on AudioSet), which raises our attention to its learning efficiency on the audio data.
Second, AST uses a class-token (CLS) to predict labels, making it unable to predict the start and end time of events in audio samples. Most CNN-based models naturally support the frame-level localization by empirically taking the penultimate layer's output as a event presence map. This inspires us to design a module that makes every output token of an audio transformer aware of the semantic meaning of events (i.e. a token-semantic module \cite{tscam}) for supporting more audio tasks (e.g. sound event detection and localization).


In this paper, we propose HTS-AT\footnote{\href{https://github.com/RetroCirce/HTS-Audio-Transformer}{https://github.com/RetroCirce/HTS-Audio-Transformer}}, a hierarchical audio transformer with a token-semantic module for audio classification. Our contributions of HTS-AT can be listed as:
\begin{itemize}[leftmargin=.15in]
    \item HTS-AT achieves or equals SOTAs on AudioSet and ESC-50, and Speech Command V2 datasets. Moreover, the model without pretraining can still achieve the performance that is only 1\%-2\% lower than the best results.
    \item HTS-AT takes fewer parameters (31M vs. 87M), fewer GPU memories, and less training time (80 hrs vs. 600 hrs) than AST's to achieve the best performance.
    \item HTS-AT further enables the audio transformer to produce the localization results of event only with weakly-labeled data. And it achieves a better performance than the previous CNN-based model. 
\end{itemize}
\vspace{-0.5cm}

\section{Proposed Model}
\vspace{-0.1cm}
\subsection{Hierarchical Transformer with Window Attention}
A typical transformer structure consumes lots of GPU memories and training time, because the length of input tokens is too long and remains unchanged in all transformer blocks from beginning to end. As a result, the machine saves the output and its gradient of each block via large GPU memories, and spends much calculation time maintaining a large global self-attention matrix. To combat these problems, as depicted in Figure \ref{fig:model-arch}, we propose two key designs: a hierarchical transformer structure and a window attention mechanism.
\vspace{-0.3cm}
\subsubsection{Encode the Audio Spectrogram}
In the left of Figure \ref{fig:model-arch}, an audio mel-spectrogram is cut into different patch tokens with a Patch-Embed CNN of kernel size $(P \times P)$ and sent into the transformer in order. 
Different from images, the width and the height of an audio mel-spectrogram denote different information (i.e. the time and the frequency bin). And the length of time is usually much longer than that of frequency bins. 
Therefore, to better capture the relationship among frequency bins of the same time frame, we first split the mel-spectrogram into \textbf{patch windows} $w_1, w_2, ..., w_n$ and then split the patches inside each window. The order of tokens follows \textbf{time$\to$frequency$\to$window} as shown in Figure \ref{fig:model-arch}.
With this order, patches with different frequency bins at the same time frame will be organized adjacently in the input sequence.

\vspace{-0.3cm}
\subsubsection{Patch-Merge and Window Attention}
In the middle of Figure \ref{fig:model-arch}, the patch tokens are sent into several groups of transformer-encoder blocks. At the end of each group, we implement a Patch-Merge layer \cite{swintransformer} to reduce the sequence size. This merge operation is applied by first reshaping the sequence to its original 2D map $(\frac{T}{P} \times \frac{F}{P}, D)$, where $D$ is the latent state dimension. Then it merges adjacent patches as $(\frac{T}{2P} \times \frac{F}{2P}, 4D)$ and finally applies a linear layer to reduce the latent dimension to $(\frac{T}{2P} \times \frac{F}{2P}, 2D)$. As illustrated in Figure \ref{fig:model-arch}, the shape of the patch tokens is reduced by 8 times from $(\frac{T}{P} \times \frac{F}{P}, D)$ to $(\frac{T}{8P} \times \frac{F}{8P}, 8D)$ after 4 network groups, thus the GPU memory consumption is reduced exponentially after each group. 

For each transformer block inside the group, 
we adopt a window attention mechanism to reduce the calculation. As shown in different color boxes in the middle right of Figure \ref{fig:model-arch}, we first split the patch tokens (in 2D format) into non-overlapping $(M \times M)$ \textbf{attention windows} $aw_1, aw_2, ..., aw_k$. Then we only compute the attention matrix inside each $M \times M$ attention window. As a result, we have $k$ window attention (WA) matrices instead of a whole global attention (GA) matrix. The computational complexities of these two mechanisms in one transformer block for $f \times t$ audio patch tokens with the initial latent dimension $D$ are:
\begin{align}
    \textrm{GA:} \quad & {\cal O} (ftD^2 + (ft)^2D) \\
    \textrm{WA:} \quad & {\cal O} (ftD^2 + M^2ftD)
\end{align}
where the window attention reduces the second complexity term by $(\frac{ft}{M^2})$ times. For audio patch tokens in a time-frequency-window order, each window attention module will calculate the relation in a certain range of continuous frequency bins and time frames. As the network goes deeper, the Patch-Merge layer will merge adjacent windows, thus the attention relation is calculated in a larger space. In the code implementation, we use the swin transformer block with a \textbf{shifted} window attention \cite{swintransformer}, a more efficient window attention mechanism. This also helps us to use the swin transformer pretrained vision model in the experiment stage.


\vspace{-0.2cm}
\subsection{Token Semantic Module}
The existing AST uses a class-token (CLS) to predict the classification label, which limits it from further indicating the start and end times of events as realized in CNN-based models. In the final layer output, each token contains information about its corresponding time frames and frequency bins. We expect to convert tokens into activation maps for each label-class (i.e. aware of semantic meaning \cite{tscam}). For strong-label datasets, we can let the model directly calculate the loss in specific time ranges. For weakly-labeled datasets, we can leverage the transformer to locate via its strong capability to capture the relation. In HTS-AT, as shown in the right of Figure \ref{fig:model-arch}, we modify the output structure by adding a token-semantic CNN layer after the final transformer block. It has a kernel size $(3, \frac{F}{8P})$ and a padding size $(1,0)$ to integrate all frequency bins and map the channel size $8D$ into the event classes $C$. The output $(\frac{T}{8P}, C)$ is regarded as a event presence map. Finally, we average the featuremap as the final vector $(1, C)$ to compute the binary cross-entropy loss with the groundtruth labels. Apart from the localization functionality, we also expect the token-semantic module to improve the classification performance, as it considers the final output by directly grouping all tokens .

\vspace{-0.2cm}
\section{Experiments}
\vspace{-0.2cm}
In this section, we evaluate the performance of HTS-AT in four datasets: the event classification on AudioSet \cite{audioset}, ESC-50 \cite{esc50}; the keyword spotting on Speech Command V2 \cite{speechcommandsv2}; and additionally, the event detection on DESED \cite{desed}.
\linespread{1.2}
\begin{table}[t]
\centering
\resizebox{\columnwidth}{!}{
\begin{tabular}{ccccc}
\hline 

\hline
Model   &Pretrain    & \#Params.         & mAP & Ensemble-mAP  \\ 
\hline
Baseline \cite{audioset} & \ding{55}  & 2.6M   & 0.314 & -  \\
DeepRes \cite{deepres} &  \ding{55}  & 26M     & 0.392 & -  \\
PANN \cite{pann} & \ding{55} & 81M & 0.434  & - \\
PSLA$^P$ \cite{psla} & \ding{51} &  13.6M & 0.444 & 0.474  \\ 
AST \cite{ast} & \ding{55} & 87M & 0.366 & - \\ 
AST$^P$ \cite{ast} & \ding{51} & 87M & 0.459 & 0.475 (0.485\tablefootnote{AST provides a second bigger ensemble result by using models with different patch settings, which is partially comparable with our settings.}) \\ \hline
HTS-AT$^H$ & \ding{55} & 28.8M & 0.440 & - \\
HTS-AT$^{HC}$ & \ding{55} & 31M & 0.453 & -  \\
HTS-AT$^{HCP}$ & \ding{51} & 31M & \textbf{0.471} & \textbf{0.487}\\
\hline

\hline
\end{tabular}
}
\caption{The mAP results on AudioSet evaluation set.}
\label{tab:exp-audioset}
\vspace{-0.5cm}
\end{table}

\vspace{-0.3cm}
\subsection{Event Classification on AudioSet}
\vspace{-0.2cm}
\subsubsection{Dataset and Training Detail}
\vspace{-0.1cm}
The AudioSet contains over two million 10-sec audio samples labeled with 527 sound event classes. In this paper, we follow the same training pipeline in \cite{pann,psla,ast} by using the full-train set (2M samples) to train our model and evaluating it on the evaluation set (22K samples). All samples are converted to mono as 1 channel by 32kHz sampling rate. We use 1024 window size, 320 hop size, and 64 mel-bins to compute STFTs and mel-spectrograms. As a result, the shape of the mel-spectrogram is $(1024, 64)$ as we pad each 1000-frame (10-sec) sample with 24 zero-frames ($T$=1024, $F$=64). The shape of the output featuremap is $(1024, 527)$ ($C$=527). The patch size is $4 \times 4$, the patch window length is 256 frames, and the attention window size is $8 \times 8$. Since 8 is divisible by 64, the attention window in the first layer will not span two frames with a large time difference. The latent dimension size is $D$=96 and the final output latent dimension is $8D$=768, which is consistent to AST. Finally, we set 4 network groups with 2, 2, 6, 2 swin-transformer blocks respectively. 

We follow \cite{pann,psla} to use the balance sampler, $\alpha=0.5$ mix-up \cite{mixup}, spectrogram masking \cite{specaugment} with time-mask=128 frames and frequency-mask=16 bins, and weight averaging. The HTS-AT is implemented in Pytorch and trained via the AdamW optimizer ($\beta_1$=0.9, $\beta_2$=0.999, eps=1e-8, decay=0.05) with a batch size of 128 ($32 \times 4$) in 4 NVIDIA Tesla V-100 GPUs. We apply a warm-up schedule by setting the learning rate as 0.05, 0.1, 0.2 in the first three epochs, then the learning rate is halved every ten epochs until it returns to 0.05. We use the mean average precision (mAP) to evaluate the classification performance.

\linespread{1.1}
\begin{table}[t]
\centering
\resizebox{\columnwidth}{!}{
\begin{tabular}{|cc|cc|}
\hline
Model     & ESC-50 Acc.(\%)   & Model  & SCV2 Acc.(\%)      \\ \hline
PANN \cite{pann}     & 90.5              & RES-15 \cite{res15} & 97.0               \\
AST \cite{ast}  & 95.6 $\pm$ 0.4          & AST \cite{ast}  & \textbf{98.1 $\pm$ 0.05} \\
ERANN \cite{erann} & 96.1              & KWT-2 \cite{kwt} & 97.3 $\pm$ 0.03          \\
HTS-AT    & \textbf{97.0 $\pm$ 0.2} & HTS-AT & \textbf{98.0 $\pm$ 0.03} \\ \hline
\end{tabular}}
\caption{The accuracy score results on ESC-50 dataset and Speech Command V2 (SCV2).}
\label{tab:exp-esc-scv2}
\vspace{-0.5cm}
\end{table}

\begin{table*}[t]
\centering
\resizebox{\textwidth}{!}{
\begin{tabular}{c|cccccccccc|c}
\hline

\hline
Model   &  Alarm & Blender & Cat & Dishes & Dog & Shaver & Frying & Water & Speech & Cleaner & Average \\ \hline
PANN \cite{pann} & 34.3 & 42.4 & 36.3 & 17.6 & 35.8 & 23.8 & 9.3 & 30.6 & 69.7 & 51.0 & 35.1 \\ 
HTS-AT &  \textbf{48.6} & 52.9 & 67.7 & 25.0 & 48.0 & \textbf{42.9} & 60.3 & 43.0 & 46.8 & 49.1 & 48.4 \\
HTS-AT - Ensemble & 47.5 & 55.1 & \textbf{72.4} & 30.9 & \textbf{49.7} & 41.9 & \textbf{63.2} & \textbf{44.3} & 51.3 & 50.6 & 50.7 \\ \hline
Zheng et al.* \cite{dcase2021} & 41.4 & 54.1 & \textbf{72.4} & 29.4 & 47.8 & \textbf{61.01} & 49.2 & 33.7 & \textbf{69.5} & 65.5 & \textbf{52.4} \\
Kim et al.* \cite{dcase2021} & 34.7 & \textbf{59.8} & 71.6 & 40.4 & 47.3 & 26.2 & 61.8 & 32.8 & 64.9 & \textbf{66.7} & 50.6 \\
Lu et al.* \cite{dcase2021}  & 37.1 & 41.4 & 62.5 & \textbf{40.6} & 39.7 & 46.5 & 46.5 & 34.5 & 54.5 & 46.9 & 45.0 \\ 
\hline

\hline
\end{tabular}
}
\caption{The event-based F1-scores of each class on the DESED test set. Models with \textbf{*} are from DCASE 2021 \cite{dcase2021}, which are partial references since they use extra training data and are evaluated on DESED test set and its another private subset.}
\label{tab:exp-desed}
\vspace{-0.5cm}
\end{table*}

\vspace{-0.4cm}
\subsubsection{Experimental Results} In Table \ref{tab:exp-audioset}, we compare our HTS-AT with different benchmark models and three self-ablated variations: (1) $H$: only hierarchical structure; (2) $HC$: with hierarchical structure and token-semantic module; and (3) $HCP$: (2) with pretrained vision model (the full setting). Our best setting achieves a new SOTA mAP 0.471 in a single model as a large increment from 0.459 by AST. We also ensemble six HTS-ATs with different training random seeds in the same settings to achieve the mAP as 0.487, and outperforms AST's 0.475 and 0.485. We analyze our results in two facets.
\vspace{-0.3cm}
\paragraph*{Token Semantic Module and Pretraining} 
PSLA, AST and HTS-AT adopt the ImageNet-pretrained model, where PSLA uses the pretrained EfficientNet \cite{effinet}, AST uses DeiT \cite{deit}, and our HTS-AT uses the swin-transformer in Swin-T/C24 setting\footnote{https://github.com/microsoft/Swin-Transformer} for $256 \times 256$ images ($256 \times 256=1024 \times 64$ as we could transfer the same size weights). We can see that the unpretrained single HTS-AT can achieve an mAP as 0.440. It is improved to 0.453 by the addition of token semantic module, 1.8\% lower than 0.471. Finally the pretrained HTS-AT achieves the new best mAP as 0.471. However, the unpretrained single AST only reflects 0.366, 9.3\% lower than 0.459. These indicate that: (1) the pretrained model definitely improves the performance by building a solid prior on pattern recognition; and (2) HTS-AT shows a far better scalability to different hyperparameters than AST, since its unpretrained model can still achieve the third best performance.


\paragraph*{Parameter Size and Training Time}
When comparing the parameter size of each model, the AST has 87M parameters. And HTS-AT is more lightweight with 31M parameters, which is even compatible with CNN-based models. As for the estimated training time, PANN takes about $72$ hours to converge and HST-AT takes about $20 \times 4=80$ hours in V-100 GPUs; and AST takes about $150 \times 4 = 600$ hours in 4 TITAN RTX GPUs\footnote{We make memories not exceed 12GB in V-100 in line with TITAN RTX.}. The speed improvement corresponds to the less calculation and GPU memory consumption of HTS-AT, as we could feed 128 samples instead of only 12 samples in AST per batch. Therefore, we conclude that HTS-AT consumes less training time and has fewer parameters than AST's, which is more efficient. 



\vspace{-0.2cm}
\subsection{Evaluations on ESC-50 and Speech Command V2}
\vspace{-0.1cm}
\subsubsection{Dataset and Training Detail}
The ESC-50 dataset contains 2000 5-sec audio samples labeled with 50 environmental sound classes in 5 folds. We train the model for 5 times by selecting 4-fold (1600 samples) as training set and the left 1-fold (400 samples) as test set. And we repeat this experiment 3 times with different random seeds to get the mean performance and deviation. The Speech Command V2 contains 105,829 1-sec spoken word clips labled with 35 common word classes. It contains 84843, 9981, and 11005 clips for training, validation and evaluation. Similarly, we train our HTS-AT for 3 times to obtain the prediction results. We use the mean accuracy score (acc) for the evaluation on both datasets. For the data processing, we resample the ESC-50 samples into 32kHz and the Speech Command clips 16kHz. And we follow the same setting of AudioSet to train the model.

\vspace{-0.4cm}
\subsubsection{Experimental Results}
\vspace{-0.1cm}
We use our best AudioSet-pretrained HTS-AT to train on these two dataset respectively and compare it with benchmark models (also in AudioSet or extra data pretraining). Since 1-sec and 5-sec does not take the full 10-sec input trained on AudioSet, we repeat the 1-sec and 5-sec by 10 and 2 times to make it 10-sec. As shown in Table \ref{tab:exp-esc-scv2}, the results shows that our HTS-AT achieves a new SOTA as 97.0\% on ESC-50 dataset and equals the SOTA 98.0\% on Speech Command V2. Our deviations are relatively smaller than AST's, indicating that HTS-AT is more stable after convergence.

\vspace{-0.3cm}
\subsection{Localization Performance on DESED}
We additionally evaluate HTS-AT's capability to localize the sound event as start and end time in given audio samples. We use the DESED test set \cite{desed}, which contains 692 10-sec test audio samples in 10 classes with the strong labels. We mainly compare our HTS-AT with PANN. We do not include AST and PSLA since AST does not directly support the event localization and the PSLA's code is not published. We also compare it partially with models in DCASE 2021 \cite{dcase2021}, nevertheless they use extra training data and are evaluated on DESED test set and its another private subset. We use the event-based F1-score on each class as the evaluation metric, implemented by a Python library \texttt{psds\_eval}\footnote{https://github.com/audioanalytic/psds\_eval}. 

The F1-scores on all 10 classes in the DESED by different models are shown in Table \ref{tab:exp-desed}. We find that HTS-AT achieves better F1-scores on 8 classes and a better average F1-score 50.7\% than PANN. When compared among leaderboard models, our model still achieves some highest scores of certain classes. However, the F1-scores on Speech and Cleaner are relatively low, indicating that there are still some improvements for a better localization performance. From the above experiments, we can conclude that HTS-AT is able to produce the specific localization output via the token-semantic module, which extends the functionality of the audio transformer. 

\vspace{-0.4cm}
\section{Conclusion and Future Work}
\vspace{-0.2cm}
In this paper, we propose HTS-AT: a hierarchical token-semantic transformer for audio classification. It achieves a new SOTA on multiple datasets of different audio classification scenarios. Furthermore, the token-semantic module enables HTS-AT to locate the events start and end time. Experiments show that HTS-AT is a high performance, high scalability, and lightweight audio transformer. In the future, we notice that a partial strong labeled subset of AudioSet has just been released \cite{audioset-strong}, we decide to conduct a detail localization training and evaluation work by HTS-AT to further explore its potential. Combining the audio classification model into more downstreaming tasks \cite{ke-sep, ke-cmg} is also considered a future work.

\bibliographystyle{IEEEbib}
\bibliography{refs}

\end{document}